\def   \ni {\noindent}
\def   \bsk {\vskip 15truept}

\documentstyle[epsfig]{article}
\begin{document}

\font\affiliation=cmssi10
\font\author=cmss10
\font\caption=cmr8
\font\references=cmr8
\font\title=cmssbx10 scaled\magstep2
\def\ref{\par\noindent\hangindent 15pt}
\null

\title{\ni Spectral properties of the Narrow-Line 
             Seyfert 1 Quasar PG1211+143
 }
                                
\bsk \bsk
\author{\ni A.~Janiuk $^{1}$, B.~Czerny $^{1}$, G.M.~Madejski $^{2}$}

\bsk
\affiliation{1) N. Copernicus Astronomical Centre,  Bartycka 18, 00-716, Warsaw, Poland\\

2) Stanford Linear Accelerator Center, 2575 Sand Hill Road, Menlo Park, CA 94025, USA
}       
\bsk
\baselineskip = 12pt

\abstract{\ni

We propose a theoretical model to explain the spectrum of the quasar PG1211+143
emitted in the Optical/X-ray bands.
In particular, we suggest that the inner accretion disk may develop a warm, optically thick skin, which produces a profound emission feature observed in the soft X-ray band. This is well modelled with the Comptonized black body emission.

The same warm, mildly ionized medium may also be responsible for the hard X-ray reflection and the presence of the iron $K_{\alpha}$ line. However, in our model it still remains an open question, whether the seed photons for Comptonization come from the cold accretion disc or from the hotter plasma. High resolution spectroscopy available through the Chandra and XMM data may provide now an independent test of the physical conditions in the Comptonizing and reflecting warm skin. 

}                                                    
\bsk
              
\section{Introduction}

The Narrow Line quasar PG1211+143 is a very good 
candidate to study the broad band emission and variability properties
due to a rich sample of observational data available for this object.
The source exhibits particularly strong hard optical spectrum and 
profound steep soft X--ray emission (Elvis, Wilkes, \& Tananbaum 1985).
It has been frequently argued that the optical/UV and soft X--ray
emission form a single Big Blue Bump component which extends across the 
unobserved XUV band (Bechtold et al. 1987), and strongly dominates the 
bolometric luminosity.
However, the nature of the soft X--ray emission below $\sim 1$ keV
is still not determined unambiguously, because any
interpretation relies on extrapolating the spectrum over a
decade in frequency through the unobserved XUV band.

Detailed analysis of the broad band spectra of Seyfert 1 galaxy NGC~5548
(Magdziarz et al. 1998)
led to the conclusion that the optical/UV spectrum is well modeled
by an accretion disk, and the hard X--rays are reproduced by standard 
thermal Comptonization.  However, an additional component is needed to 
model the
soft X--ray source: the most viable candidate here is another, optically
thick Comptonizing medium possibly associated with the transition between the
disk and the hot plasma.

Here we investigate if a similar model also applies to PG1211+143,
and if this model is unique.
We have analyzed the data form the ROSAT setellite, 
which span the period 1991-1993, ASCA observation form June 1993 and RXTE
observation form August 1997.
The ROSAT and ASCA spectral data were reduced
using the standard software package, where for ASCA, we used data
from all four detectors (SIS0/1 and GIS2/3) fitted simultaneously.
For RXTE observation we use the spectrum derived from the summed
Proportional Counter Array (PCA) data
analyzed in a standard manner (including the background subtraction),
using the {\sc ftool} script {\sc rex}.
All analysis of the spectral X--ray data was done using the XSPEC
software package, version 10.0 (Arnaud 1996). The details are given in Janiuk
et al. 2001.

\bsk

\section{Results of the spectral analysis}

The spectral fitting best model results are given in Tables 1 
and  2. 
For the  ASCA/ROSAT we supplement the {\sc comptt} component with an 
underlying hard X--ray power law component characterized by its photon index,
 and assume fixed Galactic absorption with $N_H = 2.8 \times 10^{20}$
 cm$^{-2}$. 
In case of RXTE data we adopted the composite model consisting of a primary soft 
component (described as a Comptonized black body), power law continuum, 
and the Compton -- reflected spectrum with the iron $K\alpha$ line 
included (\. Zycki, Done \& Smith 1997;  1998).
 
The reprocessed spectrum is parameterized
by the ionization parameter $\xi=L_{X}/n_{e} r^{2}$
and the reflection amplitude $R=\Omega/2 \pi$.
The outer disk radius, as before, was fixed at $10^4 R_g$ and the power law index, describing the incident flux as a function of radius was fixed  at
the value -3.0.  However, the inclination angle and the inner radius of 
the reflecting disk surface were allowed to vary.  
The iron line energy is not a free
parameter, but it depends on the ionization state of the reflecting
medium.  Its profile is connected with the assumed geometrical
parameters: inner radius of accretion disc and viewing angle.

The amplitude of reflection is consistent with 1.0,
and the reflecting area extends down to the marginally stable orbit
which means that the disk is probably not disrupted. However, its surface
(at least in the inner parts dominating the reflection) is considerably
ionized ($\xi \sim 500$), and the neutral reflection is excluded at a
$90 \%$ confidence level.  For such a case, the resulting iron line energy
corresponds to the domination of Fe XXII  ions and its rest frame
energy is $ E \sim 6.7$ keV, while the line equivalent width is $EW \sim 150$ 
eV.
The inclination angle of the disk is low, i.e. we see the disk essentially
face on.

\begin{figure}
\centerline{\psfig{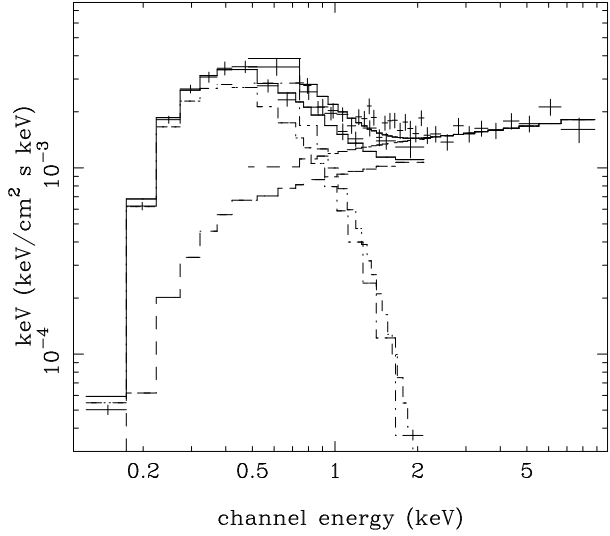}}
\caption{FIGURE 1. The model of the combined ROSAT/ASCA data (see Table 1).
\label{fig:ap1}}
\end{figure}


\begin{table}
\begin{center}
\caption{Results of simultaneous spectral fitting to ASCA and
ROSAT data}
    \renewcommand{\arraystretch}{1.2}
    \begin{tabular}[h]{lcccr}
      \hline
$\Gamma$ & $kT_{soft}$ [keV]& $kT$ [keV] & $\tau$ & $\chi^{2}_{\nu}$ \\
      \hline
$1.88 \pm 0.05$  & $<0.02$ & $0.15 \pm 0.01$ & $21 \pm 2$ & 1.18\\
      \hline \\
      \end{tabular}
    \label{tab:fits}
  \end{center}
\end{table}



\begin{figure}
\centerline{\psfig{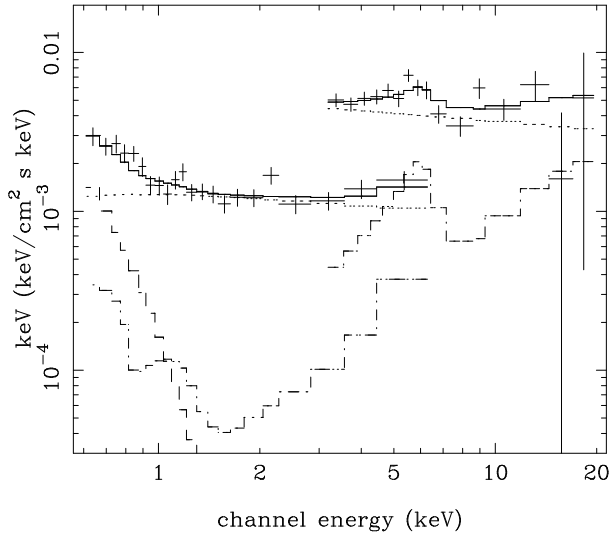}}
\caption{FIGURE 2. The model fit (continuous histogram) to the combined non-simultaneous
ASCA/XTE data (see Table 2). Dashed histogram shows the
contribution of the soft {\sc comptt} component, dotted-dashed histogram - the
reflection component and dots mark the hard X-ray power law.
\label{fig:ap2}}
\end{figure}



\begin{table}
\begin{center}
\caption{Results of simultaneous fitting to ASCA and RXTE data}
    \renewcommand{\arraystretch}{1.2}
    \begin{tabular}[h]{lcccccccr}
      \hline
$\Gamma$ &  $kT_{soft}$ [keV] & $kT$ [keV] & $\tau$ & $R$ & $\xi$ & $\cos i$ &
 $R_{in}$ [$R_{g}$] & $\chi^{2}_{\nu}$ \\

      \hline
$2.18 \pm 0.03$  & $<0.03$ & $0.13 \pm 0.05$ & $12 \pm 
5$ &
$0.85 ^{+0.65}_{-0.55} $  & $500^{+600}_{-450}$ & $0.9 \pm 0.1 $ &
$6.0 2^{+4.0}_{-0.0}$ & 1.04 \\

      \hline \\
      \end{tabular}
    \label{tab:fits_xte}
  \end{center}
\end{table}



\section{The broad band spectrum}

For the purpose of a broad band, Optical/UV/X-ray model, we add a new 
component, which essentially
has a form of standard accretion disk of Shakura \& Sunyaev (1973) 
around a non-rotating black hole.  We assume that the disk, whether 
it is or it is not covered by the ionized skin, radiates locally as 
a black body.

We assume that the emission of the standard cold disk is only
detectable from radii larger than certain $R_{ion}$.  The
gravitational energy dissipated above $R_{ion}$ is directly
re-emitted by the disk, while the energy dissipated below $R_{ion}$
has to provide the energy source for hard X--ray emission and
Comptonizing medium.
The disk parameters are therefore: the black hole mass $M$, the mass
accretion rate $\dot M$, and $R_{ion}$. We assume a face-on view
of the accretion disk, and the distance to the source calculated assuming
$H_o = 75$ km s$^{-1}$Mpc$^{-1}$.

\begin{figure}
\centerline{\psfig{file=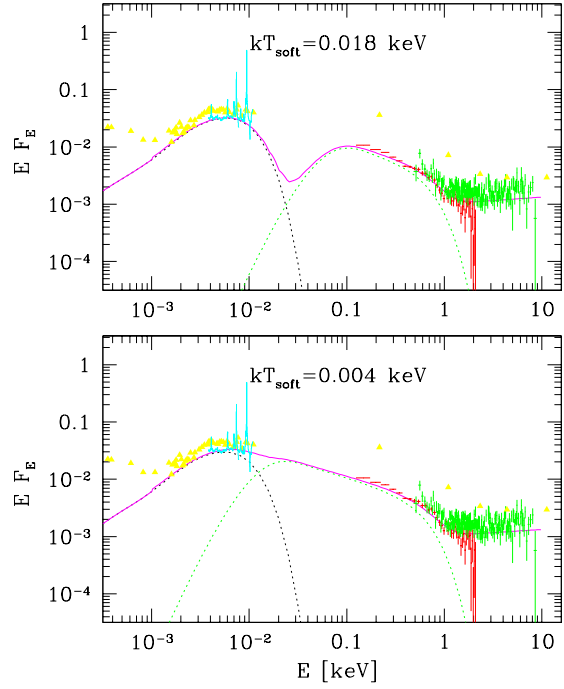, width=10cm}}
\caption{FIGURE 3. The broad band spectrum of PG1211+143 in the IR - X range.  
In the lower panel we show the model  calculated 
assuming that the accretion disk provides seed photons for Comptonization by 
the hot plasma. In the upper panel we show the model which assumes that 
the soft photons are emitted in the hotter plasma.  
The solid triangles are data points from Elvis et al. (1994). 
The UV spectrum with emission lines  are the HST data (Bechtold \& Dobrzycki 2000, private communication) and the crosses are the X--ray data from ASCA and ROSAT.
The Comptonized disk  model was fitted to the X--ray data.  
In both cases, absorption in soft X--rays is  neglected in the plot, 
and for the modeling purposes, the X--ray data were corrected for absorption.
Optical/UV spectrum was modeled with a standard disk.  
Thick solid line is the sum of all model components.
\label{spec2}}
\end{figure}


In the upper panel of the Figure ~\ref{spec2} we plot the broad band
spectrum of PG1211+143, constructed for the case of the maximum value of
temperature of the seed photons undergoing Comptonization.
In this case there is a clear gap between the UV and soft X--ray band,
which would be seen as a local minimum in the continuum around
$2\times 10^{-2}$ keV.   
The bolometric luminosity measured between 10 000 \AA~ and 100 keV (assuming
an extrapolation of the measured hard X--ray power law) is
$1.5 \times 10^{45}$ erg s$^{-1}$.
The mass of the black hole is $2.2 \times 10^8 M_{\odot}$, the
accretion rate is 0.4$M_{\odot}$ yr$^{-1}$, and $R_{ion} = 28 R_g$.
$L/L_{Edd}$ ratio in this solution is 0.05. About
50 \% of the energy is dissipated in the black body component of the
disk.

In the lower panel of Figure ~\ref{spec2} we show the broad band
spectrum determined under the assumption that cold disk itself
provides the seed photons for Comptonization, i.e. the soft photon
temperature is given by the effective temperature of the inner disk.
Such a spectrum forms a continuous Big Blue Bump extending from the
optical/UV to soft X--ray band. 
The bolometric luminosity of the source is
in this case higher, $1.9 \times 10^{45}$ erg s$^{-1}$.
The mass of the black hole is $1.5 \times 10^8$ $M_{\odot}$, the
accretion rate is 0.54 $M_{\odot}$ yr$^{-1}$, and $ R_{ion} = 42 R_g$.
$L/L_{Edd}$ ratio in this solution is equal to 0.10.
About 30 \% of the energy is directly emitted by a
black body outer disk, while 60 \% is in the Comptonized
component and $\sim 10$ \% in the hard X--ray power law.

Both fits presented above reproduce the broad band spectral
data well.  We note that all intermediate solutions are also
possible and at a level of pure data analysis, we cannot reject
any of the above possibilities.

\section{Conclusions}

\begin{figure}
\centerline{\psfig{file=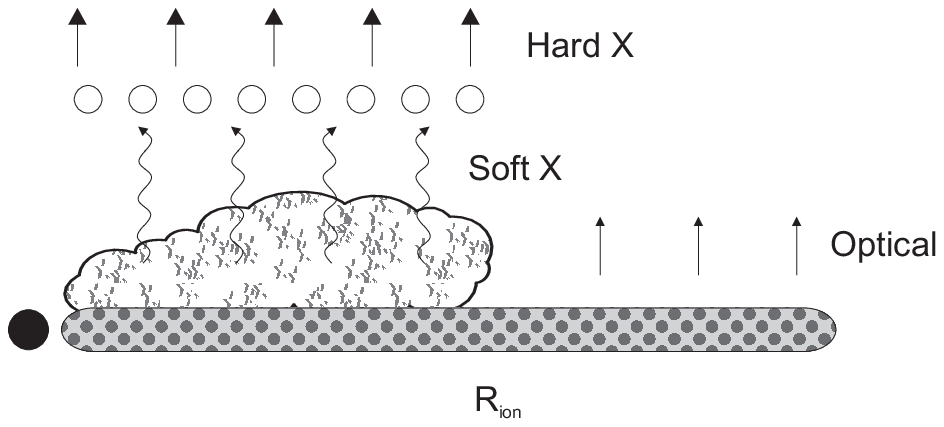, width=10cm}}
\caption{FIGURE 3. The geometry of the accretion flow in PG1211+143 consistent with the
spectral model presented in Fig. \ref{spec2}. The optical flux is
emitted by the cold accretion disk ($T \sim 10^{4}$ K). The disk 
is the source of seed photons for the hot Comptonizing cloud 
($T \sim 10^{6}$ K, $\tau \sim 20$), which extends below the 
transition radius $R_{ion}$. The hard X--ray flux is emitted by the 
hot flare region ($T \sim 10^{9}$ K) and are partially reflected by the 
cloud ($\xi \sim 500$, $\Omega/2\pi \sim 1$).  
\label{fig:geom}}
\end{figure}

Our spectral analysis requires the existence of a number of various 
plasma zones: (1) very hot plasma, responsible
for the hard X--ray emission, (2) moderately hot, moderately dense 
plasma responsible for Comptonization, (3) moderately hot 
plasma responsible for reflection and (4) a cold, dense 
disk. It is natural to pose the question whether the components (2) and
(3) are actually the same medium but parameterized differently in the
spectral fits.
Such interpretation clearly simplifies the geometry of the flow. The
accreting stream consists simply of the outer black body disk and an 
inner disk which develops very optically thick ($\tau \sim 10$), warm skin 
($T \sim 10^6 $K). A small fraction of energy is released in the form 
of hard X--ray emission above this skin.  A possible schematic view of 
the flow is shown in Figure~\ref{fig:geom}.

\bsk
\baselineskip = 12pt

{\references \ni REFERENCES
\bsk
\ref Bechtold J., Czerny B., Elvis M., Fabbiano G., Green R.F., 1987, ApJ, 314, 699

\ref Elvis M. Wilkes B., McDowell J.C., Green, R. F., Bechtold J.,
 Willner S. P., Oey M. S., Polomski E., Cutri R., 1994, ApJS, 95, 1.

\ref Janiuk A., Czerny B., Madejski G.M., 2001, ApJ, 557, 408

\ref Magdziarz P., Blaes O.M., Zdziarski A.A., Johnson
W.N., Smith D.A., 1998, MNRAS, 301, 179 
}
\end{document}